\documentclass[12pt]{article}
\usepackage{bm, color} 
\usepackage{amssymb,amsfonts,slashed,amsthm,amsmath,graphicx, soul}
\usepackage[caption=false]{subfig}
\pdfoutput=1
\usepackage{amssymb}
\usepackage{amsmath}
\usepackage{amsfonts}
\usepackage{graphicx}
\usepackage{mathrsfs}
\usepackage{comment}
\usepackage{cite}

\newcommand{\Slash}[1]{{\ooalign{\hfil#1\hfil\crcr\raise.167ex\hbox{/}}}}

\newcommand{\beq}{\begin{equation}}  \newcommand{\eeq}{\end{equation}}
\newcommand{\bef}{\begin{figure}}  \newcommand{\eef}{\end{figure}}
\newcommand{\bec}{\begin{center}}  \newcommand{\eec}{\end{center}}
\newcommand{\non}{\nonumber}  
\newcommand{\laq}[1]{\label{eq:#1}}  

\newcommand{\Eq}[1]{Eq.(\ref{eq:#1})}
\newcommand{\Eqs}[1]{Eqs.(\ref{eq:#1})}
\newcommand{\eq}[1]{(\ref{eq:#1})}
\newcommand{\Sec}[1]{Sec.\ref{chap:#1}}

\newcommand{\lac}[1]{\label{chap:#1}}

\def\({\left(}
\def\){\right)}

\def\O{\mathcal{O}}

\newcommand{\AND}{~{\rm and}~}

\newcommand{\GEV}{ {\rm \, GeV} }
\newcommand{\TEV}{ {\rm \, TeV} }

\def\a{\alpha}
\def\b{\beta}

\def\l{\lambda}
\def\m{\mu}

\def\s{\sigma}

\def\D{\Delta}

\def\tl{\tilde}
\def\*{\dagger}
\def\acent{2.51}
\def\adev{0.59}

\begin{document}
%\begin{titlepage}
\begin{center}

\hfill TU-1122

\vspace{1cm}

{\Large\bf Muon $g-2$ Anomaly in Anomaly Mediation}
\vspace{1.5cm}

Wen Yin
\vspace{1.5cm}

{Department of Physics, Tohoku University,  
Sendai, Miyagi 980-8578, Japan \\} 

\vspace{1.5cm}
\vspace{1.5cm}
\date{\today $\vphantom{\bigg|_{\bigg|}^|}$}

\abstract{
The long-standing muon $g-2$ anomaly has been confirmed recently at the Fermilab. The combined discrepancy from Fermilab and Brookhaven results shows a difference from the 
theory at a significance of 4.2 $\s$.  
In addition, the LHC has updated the lower mass bound of a pure wino. 
In this letter, we study to what extent the $g-2$ can be explained in anomaly mediation scenarios, where the pure wino is the dominant dark matter component. 
To this end, we derive some model-independent constraints on the particle spectra and $g-2$. 
We find that the $g-2$ explanation at the 1$\s$ level is driven into a corner if the higgsino threshold correction is suppressed.
On the contrary, if the threshold correction is sizable, the $g-2$ can be explained. 
In the whole viable parameter region, the gluino mass is at most $2-4\TEV $, the bino mass is at most $2\TEV$, and the wino dark matter mass is at most $1-2\TEV$. 
If the muon $g-2$ anomaly is explained in the anomaly mediation scenarios, colliders and indirect search for the dark matter may find further pieces of evidence
 in the near future. Possible UV models for the large threshold corrections are discussed. 
 }

\end{center}
\clearpage

\setcounter{page}{1}
\setcounter{footnote}{0}

%\end{titlepage}
\setcounter{footnote}{0}
\section{Introduction}

Recently Fermilab has confirmed the long-standing discrepancy of 
the muon anomalous magnetic moment ($g-2$) between the measurement at Brookhaven National Lab and 
the Standard Model (SM) prediction~\cite{Fermilab,PhysRevLett.126.141801, Bennett:2006fi}. 
The combined discrepancy is found to be
\beq
\laq{g2m}
\D a_\mu = a_\mu^{\rm EXP} - a_\mu^{\rm SM} =(\acent \pm \adev) \times 10^{-9},
\eeq
where 
$a_\mu^{\rm EXP}$ is the experimental value \cite{Fermilab, PhysRevLett.126.141801, Bennett:2006fi}~(See also Refs.~\cite{Bennett:2006fi,Roberts:2010cj, Davier:2017zfy,Keshavarzi:2018mgv,Borsanyi:2020mff, Aoyama:2020ynm, Chao:2021tvp}).
The deviation is at a significance of $4.2 \sigma$.  If we adopt the R-ratio analysis in \cite{Keshavarzi:2019abf} the significance rises to $4.5\s$ level.
This is an important message that there is a beyond SM (BSM) particle coupling to muon and whether muon $g-2$ can be explained will become an important criterion for the BSM model-building. 
The BSM should have consistent cosmology, suppressed lepton flavor violation (LFV), and CP violation (CPV). 
The minimal supersymmetric (SUSY) extension of the standard model (MSSM) can induce the $g-2$, since the muon must be coupled to superpartners like smuons and gauginos. 
Whether a SUSY model has a safe cosmology, flavor and CP structures depends on the mediation mechanism of a SUSY breaking.

There is a simple and cosmologically safe mediation effect called 
anomaly mediation~\cite{Giudice:1998xp,Randall:1998uk}. 
This loop effect exists generically and 
 is important if the tree-level mass terms for the gauginos are suppressed. 
In particular, the tree-level gaugino mass is absent if the SUSY breaking field is charged or 
sequestered~\cite{Inoue:1991rk}. 
The resulting gaugino masses are purely induced from the anomaly mediation. The masses follow the pattern:
\begin{align}
M_{\rm wino}&=3\times 10^{-3}  c_2  (m_{3/2}+L) \non \\
M_{\rm bino}&= 10^{-2} c_1 (m_{3/2}+\frac{1}{11} L)\non \\
M_{\rm gluino}&=  3 \times 10^{-2} c_3m_{3/2} \laq{mass}
\end{align} 
with $c_i\approx 1$ representing the threshold corrections 
by integrating out scalar particles, $m_{3/2}(>0)$ is the gravitino mass, 
\beq
L= \mu \sin{2\b} \frac{m_A^2}{|\m|^2-m_A^2} \log{\frac{|\mu|^2}{m_A^2}}
\eeq
representing the Higgsino threshold correction, $\mu$ the Higgsino mass parameter, $\tan\b$ the ratio of the vacuum expectation values (VEVs) of the two higgs fields, and $m_A$ the MSSM higgs boson mass. 
In general, by taking $m_{3/2}$ and VEV to be reals, $\mu$ is a complex parameter. For simplicity and for ease to evade the CPV bounds, we will limit ourselves to the purely real case, but we will come back this point in the last section. 
when the higgsino threshold correction is negligible, $L$ vanishes.  
This spectrum is almost UV model-independent. It is only corrected slightly at the renormalization scale below the splitting scale between the sparticle and particle masses. 
 For instance, the spectrum remains intact even if the model has any multiplets in the intermediate scales.\footnote{A change of the mass spectrum can occur if the Higgs boson is a slepton and there are no Higgsino multiplets at the low energy scale~\cite{Yin:2018qcs}. However, the top Yukawa coupling is difficult to be generated due to holomorphy. 
 }

For $-3 \lesssim L/m_{3/2}\lesssim 3$, the wino is the lightest gaugino and can be the dominant dark matter component if it is lighter than the other sparticles.\footnote{Out of this range, the bino can be the LSP. It, however, over-closes the universe and is excluded.    }
Since there is no need to introduce a Polonyi field for generating 
the gaugino mass, there is no Polonyi/Moduli problem. The gravitino problem is also absent 
since  the heavy gravitino has a lifetime shorter than a second. 
On the other hand, the late-time decay of the gravitino produces the wino LSP. 
The wino dark matter abundance can be explained for a certain reheating temperature even if the thermal production is not enough~\cite{Gherghetta:1999sw,Moroi:1999zb} (see also \Sec{dis}). 

In contrast to the gaugino masses, the sfermion and higgsino mass spectra are model-dependent.  There are various simple models categorized by the SUSY spectra. 
The split SUSY has higgsino as light as gaugino while others are much heavier~\cite{ArkaniHamed:2004fb, Giudice:2004tc, ArkaniHamed:2004yi}. 
The pure gravity mediation or mini-split SUSY has all other fields much heavier than the gauginos~\cite{Ibe:2011aa, Arvanitaki:2012ps, Hall:2012zp, ArkaniHamed:2012gw}. (See also Ref.~\cite{Wells:2003tf})
By taking account of a Higgs mediation~\cite{Yamaguchi:2016oqz},
i.e. in the Higgs-anomaly mediation,   the sfermions of the first two generations are as light as gauginos while others are heavy~\cite{Yin:2016shg, Yanagida:2016kag,Yanagida:2018eho, Yanagida:2020jzy}.\footnote{
 The setup is easily realized if the fermion multiplets are sequestered from the SUSY breaking but the Higgs multiplets are not. 
 If the sfermions are pseudo-Nambu-Goldstone boson in a SUSY Non-linear sigma model, a similar tree-level condition can be obtained~\cite{Yanagida:2016kag}. However, the loop induced gaugino mass spectrum is found to be different due to the K\"{a}hler structure~\cite{Yanagida:2019evh}, and, interestingly, predicts a bino-wino coannihilation. 
If there are light moduli, the F-term contribution can affect the gaugino mass spectrum. 
 In these cases, we need a solution to the moduli problem. These models are not belong to the category of this Letter's focus. 
  }  
In all the aforementioned models, flavor violation is suppressed, cosmology is consistent, and the predicted SM Higgs boson mass can easily match the measured one. (See also other early SUSY models explaining the $g-2$~\cite{Baer:2004xx, Ibe:2012qu,Okada:2012nr,Padley:2015uma, Okada:2016wlm,Abdughani:2019wai})

In this letter, we perform a model-independent  analysis to study to what extent the anomaly mediation scenarios can explain the muon $g-2$. 
We derive  the upper limit of the $g-2$ in the anomaly mediation scenarios. 
Then we show that the $g-2$ is difficult to be explained if the higgsino threshold correction is negligible, i.e. $L\sim 0$. 
On the other hand, it can be explained if the higgsino threshold correction is sizable. 
The upper bounds of the gaugino masses are derived.

\section{Effective theory for $g-2$ in anomaly mediation scenarios}
To perform a model-independent analysis, we consider an effective theory with only gauginos, satisfying \Eq{mass}, and smuons in addition to the SM particle contents.  
We do not consider a light higgsino because the enhanced DM-nucleon coupling is strongly disfavored by the direct detection experiments~\cite{Akerib:2016vxi, Cui:2017nnn,Aprile:2018dbl}.\footnote{When higgsino is much lighter than the wino, this is another model-independent setup for the dark matter and $g-2$. For further details of this scenario,
see a recent study \cite{Chakraborti:2021kkr}.} 
The LHC data, then, sets a stringent bound on the wino LSP and thus the lepton mass:
\beq
\laq{boun}
m_{smuons}\gtrsim M_{\rm wino}\gtrsim 660 (474)\GEV,
\eeq
which is reported by ATLAS \cite{ATLAS:2021ttq} (CMS \cite{Sirunyan:2020pjd}). We will take the $660\GEV$ in the following.\footnote{This bound depends on the chargino-neutralino mass splitting. 
Although the light smuons with sizable left-right-mixing contribute to the splitting, the splitting is not generated at the one loop level and thus is negligible compared to the electromagnetic contributions. 
This pure wino bound applies to our effective theory.}
This is comparable or more stringent than the indirect detection bound (e.g.~\cite{Bhattacherjee:2014dya, Reinert:2017aga, Ando:2021jvn}). 
The wino dark matter satisfying this bound may be tested  in the future not only by the collider searches but also by the direct detection experiments. 
The   LHC bounds other than \eq{boun} are much weaker in this model. The smuon bound is almost absent 
since the wino satisfying \eq{boun} is the LSP~\cite{Aad:2019vnb,Sirunyan:2020eab}. The predicted gluino mass is almost not constrained 
 if \eq{mass} and \eq{boun} are satisfied with the wino LSP~\cite{Aad:2020aze, Sirunyan:2019ctn}.\footnote{
A tiny parameter range with large $|L|$ and small masses of the bino and wino is excluded.  
 If we introduce more light sparticles like selectron, the LHC bound may become more severe. 
We do not do this as we can easily find that the resulting upper bound of $g-2$ decreases due to the higher mass scale of the sparticles. }

Since the higgsino is heavy, the only important contribution to the $g-2 $ is from a bino-smuon loop. 
The relevant effective interacting Lagrangian is given by 
\beq
{\cal L}\approx  \sqrt{2}g_Y \bar{\l}_{\rm bino} \mu_R \tl{\mu}_R^* -\frac{g_Y}{\sqrt{2}} \bar{\l}_{\rm bino} \mu_L \tl{\mu}_L^* +h.c.
\eeq
$h$ ($\l_{\rm bino}$) being the Higgs boson (bino), and $\mu_{L,R}$ ($\tl{\mu}_{L,R}$) are left, right-handed (s)muons. 
The kinetic terms are normalized. 

The smuon has a mass mixing of 
\beq
{\cal L}_{\rm LR}\approx \frac{1}{\sqrt{2}}   M_{\rm LR}  \tl{\mu}_L^* \tl{\mu}_R h+ h.c.
\eeq
where the mixing parameter is defined as
\beq
M_{\rm LR}\equiv \frac{m_\mu}{{v(1+\D)}} \mu \tan\b.
\eeq
Here, $v\approx 174\GEV$ is the SM Higgs VEV. $\D$ represents the threshold correction to the muon Yukawa coupling, and $\D/(1+\D)$ is the fraction of the muon mass that is radiatively induced. 
In addition, we define \beq m_{\tl \mu_L}^2 \AND m_{\tl \mu_R}^2\eeq as diagonal elements of the mass squared matrix for the left-handed and right-handed smuons, respectively.

The most important bound is from the vacuum (meta) stability: 
\beq
\laq{vs}
M_{\rm LR}^2\lesssim \max{[m^2_{\tl\m_L},m^2_{\tl \mu_R}]}.
\eeq
This bound can be understood since the action for the bounce solution scales as ${(m_{\tl\m_L}^2+m^2_{\tl \mu_R})}/M_{\rm LR}^2.$
A more precise fitting formula, which we adopt in the numerical simulation, can be found in Ref.~\cite{Endo:2013lva} (see also Ref.~\cite{Wainwright:2011kj}).
For given smuon diagonal mass components, this gives the maximal left-right mixing parameter, $M_{\rm LR}.$ 
By taking the mass-insertion approximation justified when $v M_{\rm LR}\ll  \max{[m_{\tl\m_L}^2,m^2_{\tl \mu_R}]}$, we obtain~\cite{Moroi:1995yh, Endo:2013lva, Cho:2011rk,Marchetti:2008hw,Degrassi:1998es}
\begin{align}
\non
(a_\mu)_{\rm SUSY} &\simeq \left( 
1 - \frac{4\a_{\rm em} }{\pi }\log{\frac{\min{[m_{\tl{\mu}_L},m_{\tl{\mu}_L}]}}{m_\mu}}
\right) \\
%\frac{3}{5}
\laq{g2}
&\times\frac{g_Y^2}{ 16\pi^2}{ \frac{m_\mu v  M_{\rm LR} M_1 } {M_1^4}} 
 \,f\left( 
\frac{m_{\tilde{\mu}_L}^2}{M_1^2},
\frac{m_{\tilde{\mu}_R}^2}{M_1^2}
\right) . 
\end{align}
where $m_{\mu}$ is the muon mass; $\a_{\rm em}\approx 1/128$; 
$f(x,y)=  \frac{(-3 + x + y + xy)}{(x - 1)^2(y - 1)^2} + 
    \frac{2 x \log{x}}{(x-y)(x - 1)^3} -\frac{ 
    2 y \log{y}}{(x-y)(y - 1)^3}$; We have not written down the radiative corrections by the integration of the sparticles above the smuon mass scale 
    because it is model-dependent. This uncertainty will be taken account by varying $c_i$. 
On the other hand, the electromagnetic correction below the smuon mass scale has been included. 
One can see that given the smuon and bino masses, the vacuum stability bound sets an upper bound for $(a_\mu)_{\rm SUSY}$. 
We note that \Eqs{vs} and \eq{g2} only depend on a combination of $\m, \tan\b, \AND \D$ in $M_{\rm LR}$. This means that our analysis does not depend on the size of $\tan \b$, or on whether the muon mass is radiatively induced.

\begin{figure}[!t]
\begin{center}  
   \includegraphics[width=105mm]{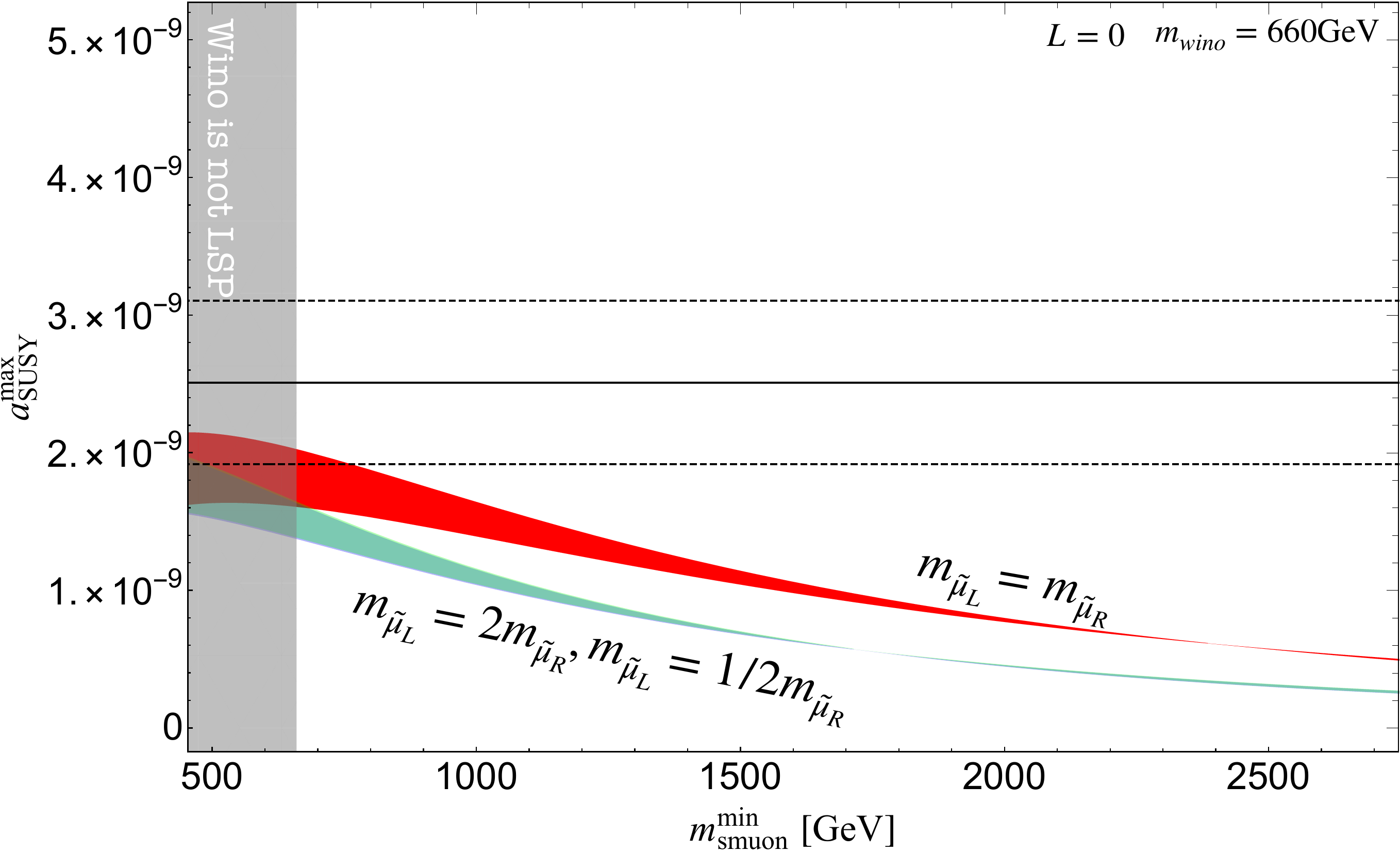}
   \includegraphics[width=105mm]{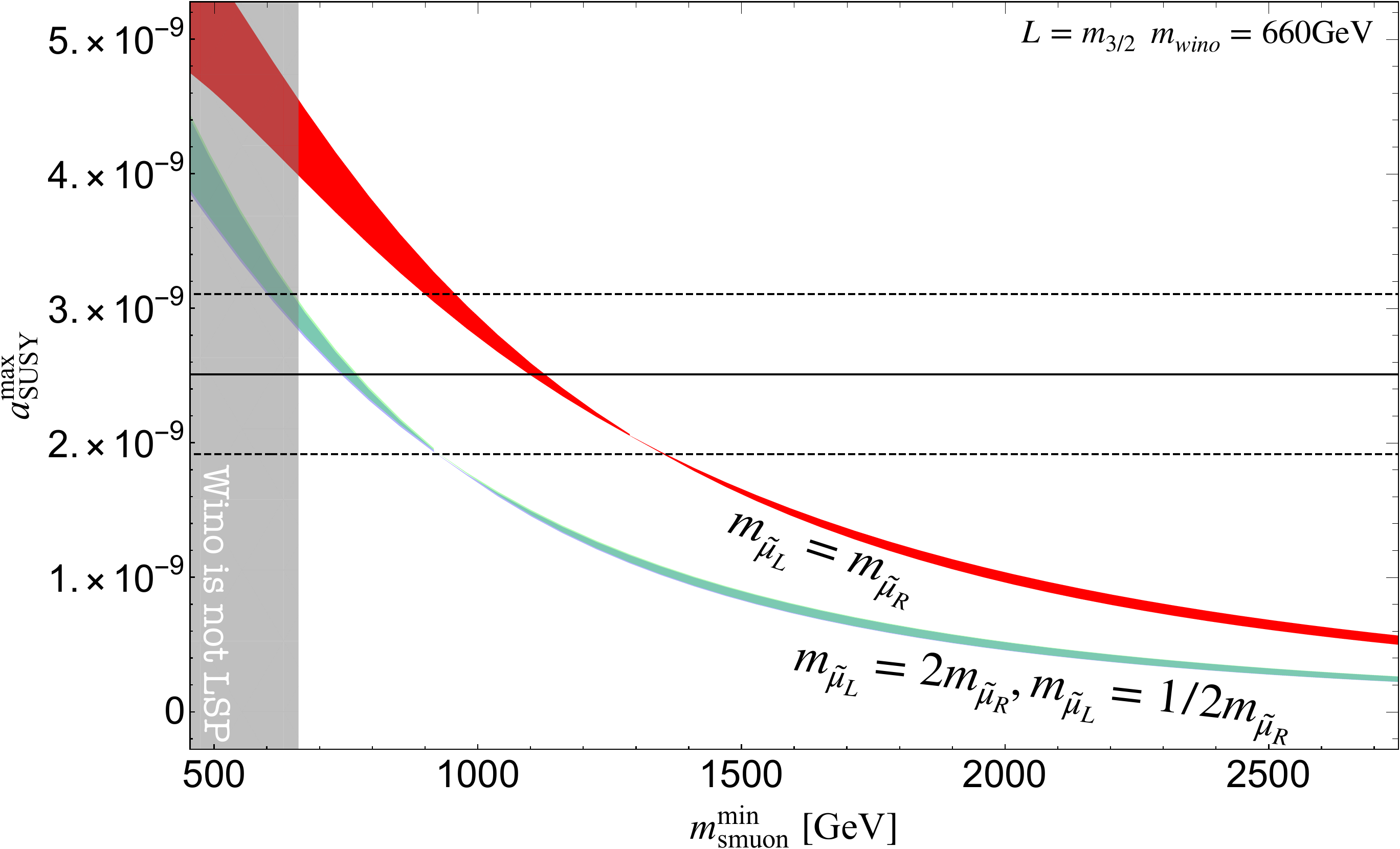}
\end{center}
\caption{The maximal SUSY contribution to the muon $g-2$ [red band] by varying the lightest smuon mass. We take the wino mass at the lower bound $M_{2}\approx 660\GEV$~\cite{ATLAS:2021ttq}. In the top (bottom) panel, $L=0$ ($L=m_{3/2}$). In the grey region $m_{\rm wino}>m^{\rm min}_{\rm smuon}$. 
The light blue band denotes the case $m_{\tl{\mu}_L}=2m_{\tl{\mu}_R}$ and $m_{\tl{\mu}_L}=0.5 m_{\tl{\mu}_R}$, which almost overlap with each other.
The 1 $\sigma $ range (central value) of the $g-2$ discrepancy is also shown by the horizontal dotted lines (solid line). 
}
\label{fig:1}
\end{figure}

In Fig.\ref{fig:1}, we show the maximized $g-2$, $a_{\rm SUSY}^{\rm max}$ [red band], evading the vacuum stability bound by varying the lightest smuon mass for  $L=0$ (left panel) and $L=m_{\rm 3/2}$ (right panel).  
We fix the wino mass to be the lowest value of $660\GEV$ from the current LHC bound. 
$a_{\rm SUSY}^{\rm max}$ corresponds to $m_{\tl{\mu}_L}=m_{\tl{\mu}_R}$ with a given lightest smuon, $m_{\rm smuon}^{\rm min}$. 
To show this, we also display a light-blue band with smuon mass splittings, $m_{\tl{\mu}_L}=2m_{\tl{\mu}_R}$ and $m_{\tl{\mu}_L}=0.5 m_{\tl{\mu}_R}$. They almost overlap. 
Thus, a smuon mass splitting leads to a smaller $a_{\rm SUSY}^{\rm max}$ than the degenerate mass case. 
As mentioned, $c_1/c_2$ is varied within $1\pm 0.05$ to take into account the model-dependent loop corrections, which give the uncertainty of the prediction. 
As a result, when $L=0$, i.e. the higgsino threshold correction is neglected, the $g-2$ can be explained at the $1\s$ level in a narrow region where $c_2/c_1\sim 0.95$ and  $M_{\rm LR}$ is close to the vacuum stability bound. We have to say that the $g-2$ explanation with $L=0$ is driven into a corner.
%This figure also represents the parameter region of the scenario. 

In the right panel, on the other hand, a case with a larger higgsino threshold correction with $L=m_{3/2}$ is shown. 
This shows that the $g-2$ at the $1\s$ level can be explained in a wider parameter range if $L= \O(m_{3/2}).$ 
This is because the bino mass slightly decreases with a given wino mass for larger $L$, and so the $g-2$ contribution is enhanced. 
As we will see soon in this case we can have a peculiar gaugino mass spectrum, and the gluino mass tends to be lighter than the usual prediction of the anomaly mediation with $L=0$. In the next section, we also discuss that the wino-bino coannihilation can take place with $L/m_{3/2}= 2\text{-}3.$

\begin{figure}[!t]
\begin{center}  
   \includegraphics[width=105mm]{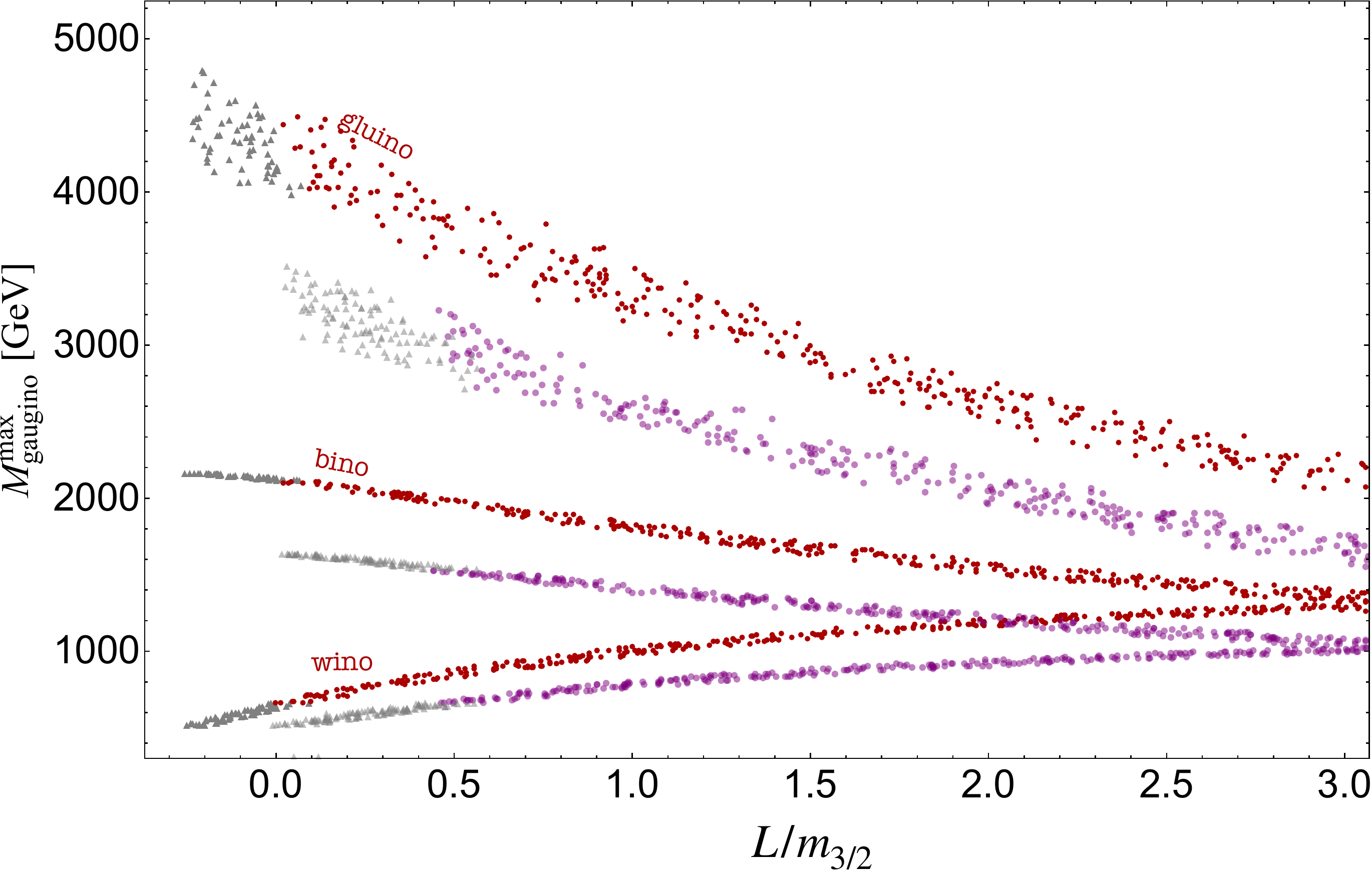}
\end{center}
\caption{The prediction of the maximized gluino bino and wino masses [the red data points]  
from top to bottom to explain the muon $g-2$ at the $1\s$ level. This corresponds to $(a_\m)_{\rm SUSY}=|\(\acent - \adev\)| \times 10^{-9}$.
The gray triangle points are excluded by the LHC (and negative $(a_\mu)_{\rm SUSY}$ if $L< 0$). 
The purple data points denote the case $(a_\m)_{\rm SUSY}=|\(\acent + \adev\)| \times 10^{-9}$ for comparison. 
}
\label{fig:2}
\end{figure}

From Fig.\ref{fig:1}, we can see that the $g-2$ is maximized with $m_{\tl{\m}_L}\approx m_{\tl{\m}_R} \approx M_{\rm LR} \approx M_{\rm wino}$ when $L$ is given. 
By using this  property, we can derive the upper bound of gaugino masses. 
Fig.\ref{fig:2} represents a scatter plot with maximized gaugino masses by varying $L$ to explain the $g-2$ at the $1\s$ level. 
We take $c_3/c_1\approx 1\pm 0.05, c_2/c_1\approx 1\pm 0.05$ at random.
The gluino, bino, and wino masses are shown by the collection of the red points from top to bottom. They are obtained by solving $10^9(a_\mu)_{\rm SUSY}= \acent - \adev$.
The gray data points in triangle are excluded due to the wino mass bound. 
We also show the case with $10^9(a_\mu)_{\rm SUSY}=\acent + \adev$ by the purple points for comparison. 

In summary, we can conclude that if the $g-2$ is explained in the anomaly mediation scenarios, it is likely that $L>0$. 
In this case, gauginos satisfy
\begin{align}
M_{\rm wino}&\lesssim 1-2\TEV \non \\
M_{\rm bino}&\lesssim 2\TEV \non\\
M_{\rm gluino}&\lesssim  2-4\TEV.
\end{align} 
The light gluino and wino masses can be tested in the LHC and future colliders~\cite{Low:2014cba, CidVidal:2018eel, Han:2020uak, Capdevilla:2021fmj, Ali:2021xlw}. 
The light bino and smuons are also predicted. Although the bino cannot be produced via electroweak process, 
we can produce it from the muon collision and then search for its decay in a muon collider~\cite{Yin:2020afe}. 
(Muon collider can test all muon $g-2$ scenarios~\cite{Capdevilla:2020qel,Buttazzo:2020eyl,Yin:2020afe, Capdevilla:2021rwo, Chen:2021rnl}.)
 In this process, we can even identify the SUSY gauge coupling as well as the bino and smuon masses~\cite{Yin:2020afe}. 
 Wino dark matter may be also searched for in direct detection experiments. 
The light gauginos as well as the light smuons with the particular mass pattern will be a smoking-gun evidence of our scenario.

\section{Discussion}

\lac{dis}
\paragraph{Wino dark matter abundance}
We have assumed the wino LSP composes the dominant dark matter component, although the thermal relic abundance of the pure wino in the mass range is smaller than the observed one.   
In fact there are two simple possibilities to realize this abundance: 
\begin{itemize}
\item{\bf Coannihilation} \\
The wino LSP mass in the range $L/m_{3/2}=2-3$ can be similar to the bino mass (see Fig.\,\ref{fig:2}). 
By increasing $L$, the thermal relic abundance of the LSP due to the coannihilation tends to increase, and it will be too much if $L\gtrsim 3m_{3/2}$ since the LSP is bino. 
Thus there must be a regime of $L$ in which the wino thermal abundance is comparable to the observed one due to the coannihilation with the bino. 
This is the case if $L\sim (2-3)m_{3/2}.$ In this scenario, the reheating temperature should be much smaller than $10^{10}\GEV$, otherwise the non-thermal component from the gravitino decay would be too large (See the following).
\item{ \bf Gravitino decay}\\
In the anomaly mediation scenario, the dark matter can be produced from the gravitino decay. The decay time is predicted to be after the freeze-out period of the wino when the wino mass is of our interest. Thus there is a non-thermal component of the wino abundance~\cite{Gherghetta:1999sw,Moroi:1999zb} 
\beq
\Omega^{\text{non-th}} h^2 \sim 0.2 \(\frac{M_{\rm wino}}{900\GEV}\) \(\frac{T_R}{3\times 10^{9}\GEV}\).
\eeq
When the thermal component is not enough, the total wino abundance can be explained by this component given a correct reheating temperature. 

\end{itemize}

\paragraph{Possible  UV models} 
So far, we found that  a large $L\sim m_{3/2}$ is favored to explain the $g-2$.
Let us discuss what kind of model can allow such a large $L$. 
One option is the pure-gravity mediation~\cite{Ibe:2011aa, Arvanitaki:2012ps, Hall:2012zp, ArkaniHamed:2012gw}, where $\tan\b= \O(1)$.  Indeed, one of the interesting predictions of the model is 
the possible large $L$. 
However we need to slightly modify the model since the smuon mass scale is much heavier than the gaugino masses in the original scenario. 
To this end, we may consider sequestering some of the lepton multiplets, including the muon multiplet, from the SUSY breaking.\footnote{One may also assume that the muon multiplet forms an $N=2$ multiplet in a high energy scale~\cite{Shimizu:2015ara, Yin:2016pkz}. 
In this case, the $N=2$ non-renormalization theorem can protect the smuons from being heavy via the SUSY breaking.}
The sequestered slepton masses are suppressed compared to the masses of other sfermions and the higgsino. 
Then we can derive $ |\mu| \sim  (1+\D)\,$PeV for  a smuon $\sim 1\TEV$. This is consistent with the Higgs boson mass and electroweak symmetry breaking in  pure-gravity mediation if $\D$ is not too large. 
Note that the charm, top, bottom, and tau multiplets may not be sequestered otherwise they are too light and the very large $\mu$-term triggers the electroweak vacuum to decay into a color/charge-breaking vacuum. 

One may also, on the other hand,  consider a large $\tan\b$ case with large $\mu$ and $m_{A}$ satisfying $m_A^2/\mu \sim \tan\b m_{3/2}.$  
In this case, we can have a Higgs mediation~\cite{Yamaguchi:2016oqz} (see also studies relevant to Higgs mediation~\cite{Endo:2019bcj, Badziak:2019gaf, Cox:2018vsv, Nagai:2020xbq}) if $m_A\lesssim \mu$. 
Then, all the lepton multiplets may be sequestered to explain the $g-2$. 
In this case the stau is heavy due to the Higgs mediation and, as a result, the stau vacuum decay problem is alleviated. 
We may also sequester the quark multiplets as in the Higgs-anomaly mediation scenarios~\cite{Yin:2016shg, Yanagida:2016kag,Yanagida:2018eho, Yanagida:2020jzy}. 
In this case, however, due to the too large $\mu$-term, the higgs mediation would induce large and negative  mass squares for the first two generation squarks. Therefore 
 the sequestering should be slightly broken to induce positive squark masses.  

By introducing the breaking of  the sequestering, 
 we may need to care about the LFV, especially the $\mu \to e \gamma$~\cite{TheMEG:2016wtm}. 
 In general, $L$'s CP phase is not aligned to that of $m_{3/2}$. Thus, we expect a CP violation. With CPV, interestingly a muon EDM can be tested in the J-PARC~\cite{Gorringe:2015cma, Crivellin:2018qmi, Abe:2019thb} (together with further confirmation of the muon $g-2$ ). 
 The gaugino masses are slightly modified due to the CP phase in $L$, which is linked to the 
muon EDM and the $g-2$. This is also a smoking-gun evidence of our scenario.
 On the other hand, the electron EDM is severely constrained~\cite{Andreev:2018ayy}. 
In this scenario, since the $\mu \tan\b$ is large, the muon (and electron) Yukawa can be easily generated radiatively, $\D \gg 1$. 
The loop-induced lepton-photon coupling and the mass basis is automatically aligned and thus the LFV is suppressed~\cite{Yin:2021yqy} (electron EDM can be also suppressed~\cite{Borzumati:1999sp,Crivellin:2010ty,Yin:2021yqy}).\footnote{One interestingly notes that if $L$ dominates over $m_{3/2}$, the scenario is CP-safe. 
The peculiar wino and bino mass relation is the prediction of this case. The suppression of the CP-violation requires a light gravitino which may be the dark matter. 
}

\section{Conclusions}
The anomaly mediation scenarios of SUSY breaking  can be easily freed from the moduli and gravitino problems. The gaugino mass relation is a renormalization invariant and thus it is the UV model-independent prediction. In this paper, we studied to what extent the muon $g-2$ anomaly can be explained within models with anomaly-induced gaugino masses. We have built an effective theory involving the gauginos and the smuons. By combining the recent LHC bound and the smuon vacuum stability bound, we found that it is hard to explain the $g-2$ if the Higgsino threshold correction is suppressed. On the other hand, if the correction is not suppressed one can still obtain a large enough $g-2$. In this case the gluino tends to be lighter than the usual case with suppressed threshold correction, and thus be produced with smaller center-of-mass energies in colliders.
The peculiar spectrum of gauginos with $L\neq 0$ and light smuons will be the smoking-gun signal in collider experiments in the future.

\section*{Acknowledgements} 
The author thanks Motoi Endo, Koichi Hamaguchi, Tsutomu T. Yanagida, and Norimi Yokozaki for discussions in different projects on muon $g-2$ within SUSY. The author also thanks 
Diego Gonzalez for carefully reading the manuscript and for useful suggestions. 
This work was supported by JSPS KAKENHI Grant-in-Aid for Scientific Research 20H05851.

\providecommand{\href}[2]{#2}\begingroup\raggedright\endgroup

\end{document}